\documentclass[aps,prb,twocolumn,groupedaddress]{revtex4}

\usepackage{mathrsfs}
\usepackage{graphicx}
\usepackage{latexsym}
\usepackage{stmaryrd}
\usepackage{amssymb}
\usepackage{wasysym}
\usepackage{longtable}
\begin{document}


\title{Vortices in a mesoscopic superconducting circular sector}


\author{Edson Sardella, Paulo Noronha Lisboa-Filho, and Andr\'e Luiz Malvezzi}
\affiliation{
Departamento de F\'{\i}sica, Faculdade de Ci\^encias \\
Universidade Estadual Paulista-UNESP \\
Caixa Postal 473, 17033-360, Bauru-SP, Brazil }


\date{\today}

\begin{abstract}
In the present paper we develop an algorithm to solve the time
dependent Ginzburg-Landau (TDGL) equations, by using the link
variables technique, for circular geometries. In addition, we
evaluate the Helmholtz and Gibbs free energy, the magnetization, and
the number of vortices. This algorithm is applied to a circular
sector. We evaluate the superconduting-normal magnetic field
transition, the magnetization, and the superconducting density.
Furthermore, we study the nucleation of giant and multi-vortex
states for that geometry.
\end{abstract}

\pacs{74.25.-q, 74.20.De, 74.78.Na}

\maketitle

\section{Introduction}
The advances in the technologies of nanofabrication in the last few
decades allowed intensive investigation efforts  in nanostructured
superconductors, both in the experimental and theoretical fronts. It
is well known that, for very confined geometries, the
superconducting-normal (SN) magnetic field transition is increased
extraordinarily. It was experimentally observed that for an Al
square very thin film with a size of few micrometers the upper
critical field $H_{c2}(T)$ can be increased up to 3.32 with the
inclusion of defects, 2.01 larger then the usual value of
$H_{c2}(T)$.\cite{berdiyorov2003,moshchalkov} Furthermore, numerical
simulations carried out in a circular wedge (see
Ref.~\onlinecite{schweigert} and references therein), have shown
that, by keeping the area of this geometry constant, the SN
transition field is a uniform increasing function with decreasing
the angular width $\Theta$ and diverges as the angle goes to zero.

Another important issue in confined geometries is the occurrence of
giant vortices. The experimental observation of giant vortex in
mesoscopic superconductor is still a controversial issue. Through
multi-small-tunnel-junction measurements in an Al thin disk film,
Kanda \emph{et al.}\cite{kanda} have argued that, as the vorticity
increases, giant vortex configuration will occur. On the other hand,
scanning SQUID microscopy on Nb thin film, both square and triangle,
cannot guarantee giant vortex configurations, at least for low
vorticity.\cite{nishiro,okayasu} Early numerical simulation of the
present authors\cite{sardella} have shown the dynamic of the
nucleation of giant and multi-vortex state before they set into an
equilibrium configuration for a square geometry.

It is well known that the phenomenology of superconductivity can be
described by the time dependent Ginzburg-Landau (TDGL)
equations.\cite{schmid} The present contribution uses the TDGL
approach to address the issues above, namely, of the nucleation of
vortices in confined geometries and the behavior of the transition
field for a deformable geometry. For this, we have chosen a circular
sector (see Fig.~\ref{fig1}), where we can arbitrarily change its
shape. To our best knowledge, the discretization of the TDGL
equations, by using the link variables technique, has been done only
in rectangular coordinates.\cite{gropp,buscaglia} So, we will extend
this algorithm to circular geometries by using polar coordinates.
Our procedure makes possible to generalize the algorithm to any
geometry. The key point in such problem is how to write the
auxiliary fields appropriately according to the system of
coordinates, making the development of the present algorithm a
specific algorithm necessary. Otherwise, the purpose of
generalization will not be achieved. We anticipate that the two
slopes of the present work will show that: \emph{(a)} as we decrease
the area of the circular sector, the transition field may increase
for large angles, but all the curves will collapse into the
asymptotic behavior $H/H_{c2}(T)=\sqrt{3}/\Theta$; and \emph{(b)}
only the confinement of vortices is not sufficient to obtain giant
vortex state, but also the geometry is very important to favor the
nucleation of such configurations. The geometry we have chosen does
not allow us to assure clearly the nucleation of giant vortex. For a
circular sector of several angular widths, ranging from $45^0$ to
$180^0$, having the same area as a disk, a square and a triangle, we
did not observe the occurrence of giant vortex as previous numerical
simulations have predicted for those geometries.\cite{baelus} In
addition, we will show that the criterion used for nucleation of
giant vortex may lead us to non conclusive pictures, at least for
the geometry under the present investigation.

The paper is outlined as follows. In Section \ref{secTDGL} we write
the TDGL equations in a gauge invariant form by using the auxiliary
field in polar coordinates. In Section~\ref{secNum} we develop the
algorithm we use to solve the TDGL equations: we define the mesh
used to discretize the TDGL for a circular sector, the discrete
variables which are evaluated in the mesh, the boundary conditions
and, finally, the important physical quantities which will be
extracted from the numerical setup are determined. In
Section~\ref{secRD} we present and discuss the results of the
numerical simulations for certain parameters of a superconducting
circular sector.

\section{The TDGL Equations} \label{secTDGL}
The properties of the superconducting
state are usually described by the complex order parameter $\psi$,
for which the absolute square value $|\psi|^2$ represents the
superfluid density, and the vector potential ${\bf A}$, which is
related to the local magnetic field as ${\bf
h}=\boldmath{\nabla}\times{\bf A}$. These quantities are determined
by the TDGL equations, which in the non-dimensional version are
given by
\begin{eqnarray}\label{TDGL}
  \frac{\partial \psi}{\partial t} &=& -{\bf D}\cdot{\bf D}\psi+(1-T)\psi(1-|\psi|^2)\;, \nonumber  \\
  \beta\frac{\partial {\bf A}}{\partial t} &=& (1-T) {\rm Re}\left [ \bar{\psi}{\bf D}\psi\right
  ]-\kappa^2\boldmath{\nabla}\times{\bf
  h}\;,
\end{eqnarray}
where $T$ is the temperature in units of the critical temperature;
lengths are in units of $\xi(0)$, the coherence length at zero
temperature, and fields in units of $H_{c2}(0)$, the upper critical
field at zero temperature; $\beta$ is the ratio between the
relaxation times of the vector potential and the order parameter;
$\kappa$ is the Ginzburg-Landau parameter which is material
dependent; the operator ${\bf D}=-i\boldmath{\nabla}-{\bf A}$; ${\rm
Re}$ indicates the real part of a complex variable and the overbar
means the complex conjugation; (for more details, see Reference
\onlinecite{sardella,gropp,buscaglia}). Here, we will neglect the
$z$-dependence on the order parameter. This is valid either if the
system is infinite along the $z$-direction or if it is a very thin
film of thickness $d\ll 1$. However, in the former case, $\kappa^2$
is replaced by an effective Ginzburg-Landau parameter
$\kappa_{eff}^2=\kappa^2/d$ (for instance, see References
\onlinecite{pearl} and \onlinecite{berdiyorov2006}). The
generalization to the a system of arbitrary thickness should not
present any difficulty.

It is convenient to introduce the auxiliary vector field ${\bf {\cal
U}}=({\cal U}_{\rho},{\cal U}_{\theta})$ in polar coordinates, which
is defined by
\begin{eqnarray}\label{auxiliaries}
  {\cal U}_{\rho}(\rho,\theta) &=& \exp\left ( -i\int_{\rho_0}^\rho\,
  A_{\rho}(\theta,\xi)\,d\xi \right )\;, \nonumber \\
  {\cal U}_{\theta}(\rho,\theta) &=& \exp\left ( -i\int_{\theta_0}^\theta\,
  A_{\theta}(\xi,\rho)\rho\,d\xi \right )\;,
\end{eqnarray}
where $(\rho_0,\theta_0)$ is an arbitrary reference point. For the sake of brevity we
omit the time dependence on the fields.

Notice that
\begin{equation}
  \frac{\partial {\cal U}_{\rho}}{\partial\rho} = -iA_{\rho}{\cal U}_{\rho}\;,\;\;\;
  \frac{1}{\rho}\frac{\partial {\cal U}_{\theta}}{\partial \theta} = -iA_{\theta}{\cal U}_{\theta}\;,
\end{equation}
and that
\begin{equation}\label{Ds}
  D_{\rho}\psi = -i\bar{\cal U}_{\rho}\frac{\partial ({\cal U}_{\rho}\psi)}
  {\partial\rho}\;,\;\;\;
  D_{\theta}\psi = -i\frac{\bar{\cal U}_{\theta}}{\rho}
  \frac{\partial ({\cal U}_{\theta}\psi)}{\partial \theta}\;.
\end{equation}

Upon using these two last equations recursively, we obtain
\begin{equation}
  D_{\rho}^2\psi =  -\bar{\cal U}_{\rho}\frac{\partial^2 ({\cal U}_{\rho}\psi)}
  {\partial\rho^2}\;,\;\;\;
  D_{\theta}^2\psi = -\frac{\bar{\cal U}_{\theta}}{\rho^2}
  \frac{\partial^2 ({\cal U}_{\theta}\psi)}
  {\partial \theta^2}\;.
\end{equation}

As a consequence, we obtain for the kinetic term in the first TDGL
equation
\begin{eqnarray}\label{Laplacian}
  {\bf D}\cdot{\bf D}\psi &=& D_{\rho}^2\psi+D_{\theta}^2\psi-\frac{i}{\rho}D_{\rho}\psi \nonumber \\
  &=&-\frac{\bar{\cal U}_{\rho}}{\rho}\frac{\partial}{\partial \rho}\left [\rho \frac{\partial ({\cal U}_{\rho}\psi)}
  {\partial\rho}\right ] -\frac{\bar{\cal U}_{\theta}}{\rho^2}
  \frac{\partial^2 ({\cal U}_{\theta}\psi)}{\partial \theta^2}\;.
\end{eqnarray}

From equations (\ref{Ds}), it can also be easily proved that
\begin{eqnarray}\label{currents}
    {\rm Re} [\bar{\psi}D_{\rho}\psi] &=& {\rm Im} \left [ \bar{\cal U}_{\rho}\bar{\psi}
    \frac{\partial ({\cal U}_{\rho}\psi)}
    {\partial\rho}\right ]\;, \nonumber \\
    {\rm Re} [\bar{\psi}D_{\theta}\psi] &=& {\rm Im} \left [ \frac{
    \bar{\cal U}_{\theta}\bar{\psi}}{\rho}
    \frac{\partial ({\cal U}_{\theta}\psi)}
    {\partial\theta}\right ]\;,
\end{eqnarray}
where ${\rm Im}$ indicates the imaginary part of a complex variable.

Finally, by using equations (\ref{Laplacian}) and (\ref{currents}),
the TDGL equations of (\ref{TDGL}) can be rewritten as
\begin{eqnarray}\label{TDGLdiffusion}
  \frac{\partial \psi}{\partial t} &=& \frac{\bar{\cal U}_{\rho}}{\rho}\frac{\partial}{\partial \rho}\left [\rho \frac{\partial ({\cal U}_{\rho}\psi)}
  {\partial\rho}\right ]+\frac{\bar{\cal U}_{\theta}}{\rho^2}
  \frac{\partial^2 ({\cal U}_{\theta}\psi)}{\partial \theta^2}\nonumber \\
  & & +(1-T)\psi(1-|\psi|^2)\;, \nonumber \\
  \beta\frac{\partial A_{\rho}}{\partial t} &=& (1-T){\rm Im} \left [
  \bar{\cal U}_{\rho}\bar{\psi}\frac{\partial ({\cal U}_{\rho}\psi)}
  {\partial\rho}\right ]-\kappa_{eff}^2\frac{1}{\rho}\frac{\partial h_z}{\partial \theta}\;,
  \nonumber \\
  \beta\frac{\partial A_{\theta}}{\partial t} &=& (1-T){\rm Im} \left [
  \frac{\bar{\cal U}_{\theta}\bar{\psi}}{\rho}\frac{\partial ({\cal U}_{\theta}\psi)}
  {\partial \theta}\right ]+\kappa_{eff}^2\frac{\partial h_z}{\partial
  \rho}\;.
\end{eqnarray}

Disregarding the non-linear term, the first TDGL equation written as
above resembles a diffusion equation, except by the fact that the
Laplacian appears with different weights. The weights depend locally
on the components of the auxiliary field ${\bf {\cal U}}$. Written
like in (\ref{TDGLdiffusion}), the TDGL equations are gauge
invariant, that is, they do not change their form under any
transformation $\psi\rightarrow \psi e^{i\chi}$, ${\bf A}\rightarrow
{\bf A}+\boldmath{\nabla}\chi$. This is a very important point for
any discretization procedure of the TDGL equations. Otherwise, we
may obtain non-physical numerical solutions.

\section{Numerical Method}\label{secNum}

\subsection{The Computational Mesh}\label{SecMesh}
\begin{figure}
  \centering
  \includegraphics[scale=0.9]{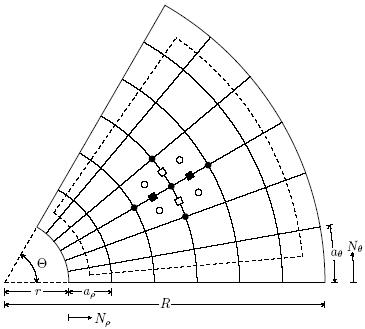}
  \caption{The computational mesh in polar coordinates used for the
  evaluation of $\psi_{i,j}$ ($\newmoon$, vertex point); $h_{z,i,j}$ and $L_{i,j}$ ($\bigcirc$, cell point);
  $A_{\rho,i,j}$ and $U_{\rho,i,j}$ ($\blacksquare$, link point); $A_{\theta,i,j}$ and
  $U_{\theta,i,j}$ ($\square$, link point). The superconducting domain
  is delimited by the dashed line $\partial\Omega_{\rm SC}$,
  and superconductor and the normal metal
  are surrounded by the solid line $\partial\Omega$. Other details of the
  Figure are described in Section \ref{SecMesh}}\label{fig1}
\end{figure}

We will discretize the TDGL equations of (\ref{TDGLdiffusion}) on a
circular sector as illustrated in Fig.~\ref{fig1}. The mesh consists
of $N_{\rho}\times N_{\theta}$ cells with size
($a_{\rho},a_{\theta})$ in polar coordinates. The circular sector
has internal radio $r$ and external $R$; $\Theta$ is its angular
width. Let $(\rho_i,\theta_j)$ be a vertex point in the mesh, where
$\rho_{i+1}=\rho_i+a_{\rho}$, $\theta_{j+1}=\theta_j+a_{\theta}$,
for all $\{1\le i \le N_{\rho},1\le j \le N_{\theta}\}$; $\rho_1=r$
and $\theta_1=0$; this particular choice for the initial value of
the angle does not imply in lost of generality since the system is
invariant under any rotation. The superconducting domain is
comprehended by $\Omega_{\rm SC}= \{\rho_1+a_{\rho}/2 < \rho <
\rho_{N_{\rho}}+a_{\rho}/2, a_{\theta}/2 < \theta <
\theta_{N_{\theta}}+a_{\theta}/2\}$. The superconducting region is
surrounded by a thin normal metal layer of width $a_{\rho}/2$ in the
radial direction. Both regions are inside the domain
$\Omega=\{\rho_1< \rho < \rho_{N_{\rho}+1},0< \theta <
\theta_{N_{\theta}+1}\}$. We denote by $\partial\Omega_{\rm SC}$ the
interface between the superconductor and the normal metal, and by
$\partial\Omega$ the normal metal-vacuum interface.

\subsection{The Discrete Variables}
Let us define the following discrete variables.

$\bullet$ The vertex points
\begin{eqnarray}
  \rho_i &=& r+(i-1)a_{\rho},\;\;\;1\le i \le N_{\rho}+1\;, \nonumber \\
  \theta_j &=& (j-1)a_{\theta},\;\;\;1\le j \le N_{\theta}+1\;,
\end{eqnarray}
The points $(\rho_{i+1/2}=\rho_i+a_\rho/2,
\theta_{i+1/2}=\theta_i+a_\theta/2)$ are called the cell points. The
points $(\rho_{i+1/2},\theta_j)$ and $(\rho_i,\theta_{j+1/2})$ are
the link points in the radial and  transversal directions
respectively (see Fig.~\ref{fig1}).

$\bullet$ The order parameter
\begin{equation}
  \psi_{i,j}=\psi(\rho_i,\theta_j)\;,
\end{equation}
for all $\{1\le i \le N_{\rho}+1\;,1\le j \le N_{\theta}+1\}$.

$\bullet$ The vector potential
\begin{equation}
  A_{\rho,i,j} = A_{\rho}(\rho_{i+1/2},\theta_j)\;,\;\;\;
  A_{\theta,i,j} = A_{\theta}(\rho_i,\theta_{j+1/2})\;,
\end{equation}
for all $\{1\le i \le N_{\rho}\;,1\le j \le N_{\theta}+1\}$, and
$\{1\le i \le N_{\rho}+1\;,1\le j \le N_{\theta}\}$, respectively.

$\bullet$ The link variables
\[
  U_{\rho,i,j} = \bar{\cal U}_{\rho}(\rho_i,\theta_j){\cal U}_{\rho}
  (\rho_{i+1},\theta_j) = \exp\left ( -ia_\rho A_{\rho,i,j} \right )\;,
\]
\begin{equation}\label{linkvariables}
  U_{\theta,i,j} = \bar{\cal U}_{\theta}(\rho_i,\theta_j){\cal U}_{\theta}
  (\rho_i,\theta_{j+1}) = \exp\left ( -i\rho_ia_\theta A_{\theta,i,j} \right
               )\;,
\end{equation}
for all $\{1\le i \le N_{\rho}\;,1\le j \le N_{\theta}+1\}$, and
$\{1\le i \le N_{\rho}+1\;,  1\le j \le N_{\theta}\}$, respectively.

$\bullet$ The local magnetic field
\begin{equation}
  h_{z,i,j} = h_z(\rho_{i+1/2},\theta_{j+1/2})\;,
\end{equation}
for all $\{1\le i \le N_{\rho}\;,1\le j \le N_{\theta}\}$.

In what follows, it will be important to define the following
discrete variable
\begin{eqnarray}\label{L}
    L_{i,j} &=& \exp\left ( -i\oint_{\partial\mathscr{D}}\,{\bf A}\cdot\,d{\bf r} \right )
    \nonumber \\
           &=& \exp\left ( -i\int_{\mathscr{D}}\,h_z\,\rho\, d\rho d\theta \right
           ) \nonumber \\
           &=& \exp\left ( -ia_{\rho}\rho_{i+1/2}a_\theta h_{z,i,j} \right )\;,
\end{eqnarray}
for all $\{1\le i \le N_{\rho}\;,1\le j \le N_{\theta}\}$, where
$\mathscr{D}$ is the domain of a unit cell limited by a closed path
$\partial\mathscr{D}$. The use of the Stoke's theorem and the
midpoint rule for numerical integration have been made. A simple
inspection of equation (\ref{L}) leads to
\begin{equation}\label{LU}
    L_{i,j}=U_{\rho,i,j}U_{\theta,i+1,j}\bar{U}_{\rho,i,j+1}\bar{U}_{\theta,i,j}\;.
\end{equation}

\subsection{The Discretization of the TDGL}
Now, we are in a position to discretize the TDGL equations. This can
be done by using the central difference approximation for the
derivatives which is second order accurate in $(a_\rho,a_\theta)$. A
tedious, however straightforward calculation, leads us to the
following discrete version of the TDGL equations of
(\ref{TDGLdiffusion})
\begin{eqnarray}\label{TDGLrecurrence1}
  \frac{\partial \psi_{i,j}}{\partial t} &=&
  \mathscr{F}_{\psi,i,j}\;,\nonumber \\
  \beta\frac{\partial A_{\rho,i,j}}{\partial t} &=& (1-T){\rm Im}\left
  [ \frac{\bar{\psi}_{i,j}U_{\rho,i,j}\psi_{i+1,j}}{a_\rho}\right ]\nonumber
  \\
  & & -\kappa_{eff}^2
  \left ( \frac{h_{z,i,j}-h_{z,i,j-1}}{\rho_{i+1/2}a_\theta}\right )\;,\nonumber \\
  \beta\frac{\partial A_{\theta,i,j}}{\partial t} &=& (1-T){\rm Im}\left
  [\frac{\bar{\psi}_{i,j}U_{\theta,i,j}\psi_{i,j+1}}{\rho_ia_\theta}\right ]\nonumber
  \\
  & & +\kappa_{eff}^2
  \left ( \frac{h_{z,i,j}-h_{z,i-1,j}}{a_\rho}\right )\;,
\end{eqnarray}
where
\begin{eqnarray}
  \mathscr{F}_{\psi,i,j} &=& \frac{1}{\rho_ia_\rho^2}\left [ \rho_{i+1/2}\left( U_{\rho,i,j}\psi_{i+1,j}-\psi_{i,j}
  \right )+\right . \nonumber \\
  & & \left . \rho_{i-1/2}\left( \bar{U}_{\rho,i-1,j}\psi_{i-1,j}-\psi_{i,j}
  \right )  \right ] \nonumber \\
  & & +\frac{U_{\theta,i,j}\psi_{i,j+1}-2\psi_{i,j}+
  \bar{U}_{\theta,i,j-1}\psi_{i,j-1}}{\rho_i^2a_\theta^2}\nonumber \\
  & & +(1-T)\psi_{i,j}(1-|\psi_{i,j}|^2)\;.
\end{eqnarray}

From the numerical point of view, it is more convenient to evaluate
the link variables rather than the vector potential. From equations
(\ref{linkvariables}), we can easily verify that
\begin{equation}\label{AUderivative}
  \frac{\partial A_{\rho,i,j}}{\partial t} = -\frac{\bar{U}_{\rho,i,j}}{ia_\rho}
  \frac{\partial U_{\rho,i,j}}{\partial t}\;,\;\;\;
  \frac{\partial A_{\theta,i,j}}{\partial t} = -\frac{\bar{U}_{\theta,i,j}}{i\rho_ia_\theta}
  \frac{\partial U_{\theta,i,j}}{\partial t}\;.
\end{equation}
In addition, from equation (\ref{L}), we can write, accurate to
second order in $(a_\rho,a_\theta)$
\begin{equation}\label{hL}
    h_{z,i,j}=\frac{{\rm Im}\left ( 1-L_{i,j}\right ) }{a_{\rho}\rho_{i+1/2}a_\theta
    }\;,
\end{equation}
where $L_{i,j}$ is given by equation (\ref{LU}). Upon introducing
equations (\ref{AUderivative}) and (\ref{hL}) into the second and
third equations of (\ref{TDGLrecurrence1}) we obtain the following
recurrence relations
\begin{equation}\label{Fs}
  \frac{\partial U_{\rho,i,j}}{\partial t} = -\frac{i}{\beta}U_{\rho,i,j}\mathscr{F}_{U_\rho,i,j}\;,\;\;
  \frac{\partial U_{\theta,i,j}}{\partial t} = -\frac{i}{\beta}U_{\theta,i,j}\mathscr{F}_{U_\theta,i,j}\;,
\end{equation}
where
\begin{eqnarray}
  \mathscr{F}_{U_\rho,i,j} &=& {\rm Im}\left [ (1-T)\bar{\psi}_{i,j}U_{\rho,i,j}\psi_{i+1,j}\right . \nonumber
  \\
  & & \left . +
  \kappa_{eff}^2\left ( \frac{L_{i,j}-L_{i,j-1}}{\rho_{i+1/2}^2a_\theta^2}\right ) \right ]\;,\nonumber \\
  \mathscr{F}_{U_\theta,i,j} &=& {\rm Im}\left [  (1-T)\bar{\psi}_{i,j}U_{\theta,i,j}\psi_{i,j+1}\right . \nonumber
  \\
  & & \left . +
  \kappa_{eff}^2\frac{\rho_i}{a_\rho^2}\left ( \frac{L_{i-1,j}}{\rho_{i-1/2}}-\frac{L_{i,j}}{\rho_{i+1/2}} \right )  \right
  ]\;. \nonumber \\
  & &
\end{eqnarray}

Finally, on using the one-step forward-difference Euler scheme with
time step $\Delta t$, we obtain the following recurrence relations
\begin{eqnarray}\label{TDGLrecurrence2}
  \psi_{i,j}(t+\Delta t) &=& \psi_{i,j}(t)+\Delta t\mathscr{F}_{\psi,i,j}(t)\;,\nonumber \\
  U_{\alpha,i,j}(t+\Delta t) &=&U_{\alpha,i,j}(t)\exp\left ( -\frac{i}{\beta}
  \mathscr{F}_{U_\alpha,i,j}(t)\Delta t \right )\;, \nonumber \\
\end{eqnarray}
where $\alpha=(\rho,\theta)$. Notice that equations
(\ref{TDGLrecurrence2}) were written in such a manner they guarantee
the link variables are unimodular functions. The first recurrence
relation run for all interior vertex points of $\Omega_{\rm SC}$,
that is, $\{ 2\le i \le N_\rho,2\le j \le N_\theta\}$; the second
ones for all link points in the interior of $\Omega$, that is, $\{
1\le i \le N_\rho,2\le j \le N_\theta\}$ for $\alpha=\rho$, and $\{
2\le i \le N_\rho,1\le j \le N_\theta\}$ for $\alpha=\theta$. At the
edge points of $\Omega$, the values of the discrete variables will
be evaluated using the boundary conditions (see next Section).

There is a severe limitation on the choice of the time step $\Delta
t$ such that the recurrence relations converge. We have learned
experimentally that the condition for stability is assured by the
following practical rule
\begin{equation}\label{Deltat}
    \Delta t \le {\rm min}\left \{
    \frac{\delta^2}{4},\frac{\delta^2\beta}{4\kappa^2}
    \right \}\;,
\end{equation}
where
\begin{equation}
    \delta^2=\frac{2}{\frac{1}{a_\rho^2}+\frac{1}{r^2a_\theta^2}}\;.
\end{equation}

Notice that the stability is controlled by the size of the smallest
unit cell. The smaller the value of $r$, the more severe the
restriction on the time step becomes. Perhaps in this case, it would
be more convenient to use either a semi or a full implicit scheme to
solve equations (\ref{Fs}), which are usually unconditionally
convergent.

\subsection{The Boundary Conditions}
Let ${\bf n}$ be a unit vector normal to the $\partial\Omega_{\rm
SC}$ interface and directed outward the domain $\Omega_{\rm SC}$. We
will assume that the normal current density vanishes at the
superconductor-normal metal interface, that is, ${\bf
D}\psi\cdot{\bf n}=0$. By using equations (\ref{Ds}), it can be
shown that the discrete implementation of this condition is as
follows
\begin{eqnarray}
  \psi_{1,j} &=& U_{\rho,1,j}\psi_{2,j}\;, \nonumber \\
  \psi_{N_\rho+1,j} &=& \bar{U}_{\rho,N_\rho,j}\psi_{N_\rho,j}\;, \nonumber \\
  \psi_{i,1} &=& U_{\theta,i,1}\psi_{i,2}\;, \nonumber \\
  \psi_{i,N_\theta+1} &=& \bar{U}_{\theta,i,N_\theta}\psi_{i,N_\theta}\;.
\end{eqnarray}
The first two equations run for all values of $\{2\le j\le
N_\theta\}$, and the second ones for all values of $\{2\le i\le
N_\rho\}$. At the corner vertex points of the domain $\Omega$ are
not necessary to run the recurrence relations
(\ref{TDGLrecurrence2}).

These last four equations update the values of the order parameter
at any vertex point at the $\partial\Omega$ interface. The values of
the link variables at this interface will be updated by using the
fact that the $z$-component of the magnetic field is continuous at
the interface $\partial\Omega_{\rm SC}$, that is,
$h_{z,1,j}=h_{z,N_\rho,j}=h_{z,i,1}=h_{z,i,N_\theta}=H$, which is
the external applied magnetic field. Consequently, from equations
(\ref{L}) and (\ref{LU}), the link variables are updated according
to
\begin{equation}
    L_{i,j}=\exp\left ( -ia_{\rho}\rho_{i+1/2}a_\theta H \right )\;,
\end{equation}
which runs for all edge points at the interface $\partial\Omega$.

\subsection{The Physical Quantities}
The topology of the superconducting state is usually illustrated by
$|\psi|^2$. This quantity can be determined from the outcome of the
recurrence relations previously derived. Other important physical
quantities used to describe the vortex state are the Gibbs free
energy, the magnetization, and the vorticity. In what follows we
will derive an expression for each of these physical quantities.
\begin{widetext}
$\bullet$ The kinetic energy
\begin{eqnarray}
  \mathscr{L}_{\rm k} &=& (1-T)\,\iint_{\Omega_{\rm SC}}\,\left [
  \left | \frac{\partial (U_\rho\psi)}{\partial \rho} \right |^2+
  \left | \frac{1}{\rho^2}\frac{\partial (U_\theta\psi)}{\partial \theta}
  \right |^2\right ]\,\rho\,d\rho d\theta\nonumber \\
  &=& (1-T)\,\sum_{i=2}^{N_\rho}\,\sum_{j=2}^{N_\theta}\,\int
  _{\theta_{j-1/2}}^{\theta_{j+1/2}}\,\int
  _{\rho_{i-1/2}}^{\rho_{i+1/2}}\,\left [
  \left | \frac{\partial (U_\rho\psi)}{\partial \rho} \right |^2+
  \left | \frac{1}{\rho^2}\frac{\partial (U_\theta\psi)}{\partial \theta}
  \right |^2\right ]\,
  \rho\,d\rho d\theta\nonumber \\
  &=& (1-T)\,\sum_{i=2}^{N_\rho}\,\sum_{j=2}^{N_\theta}\,\left \{
  \frac{1}{2\rho_ia_\rho^2}\left [ \rho_{i+1/2}\left | U_{\rho,i,j}\psi_{i+1,j}-\psi_{i,j}
  \right |^2\right . \left . +\rho_{i-1/2}\left | U_{\rho,i-1,j}\psi_{i,j}-\psi_{i-1,j}
  \right |^2\right ] \right . \nonumber \\
  & & \left .
  +\frac{1}{2\rho_i^2a_\theta^2}\left [ \left |
  U_{\theta,i,j}\psi_{i,j+1}-\psi_{i,j}
  \right |^2+\right .
  \left . \left | U_{\theta,i,j-1}\psi_{i,j}-\psi_{i,j-1}\right
  |^2
  \right ]
  \right \}a_\rho a_\theta\rho_i\;.
\end{eqnarray}
\end{widetext}
$\bullet$ The condensation energy
\begin{eqnarray}
  \mathscr{L}_{\rm c} &=& (1-T)^2\,\iint_{\Omega_{\rm SC}}\,|\psi|^2
  \left ( \frac{1}{2}|\psi|^2-1\right )\,
  \rho\,d\rho d\theta\nonumber \\
  &=& (1-T)^2\,\sum_{i=2}^{N_\rho}\,\sum_{j=2}^{N_\theta}\,\int
  _{\theta_{j-1/2}}^{\theta_{j+1/2}}\,\int
  _{\rho_{i-1/2}}^{\rho_{i+1/2}}\,|\psi|^2\times\nonumber
  \\
  & & \left ( \frac{1}{2}|\psi|^2-1\right )\,
  \rho\,d\rho d\theta\nonumber \\
  &=& (1-T)^2\,\sum_{i=2}^{N_\rho}\,\sum_{j=2}^{N_\theta}\,|\psi_{i,j}|^2\left (
  \frac{1}{2}|\psi_{i,j}|^2-1\right )a_\rho a_\theta\rho_i\;. \nonumber \\
\end{eqnarray}

$\bullet$ The field energy
\begin{eqnarray}
  \mathscr{L}_{\rm f} &=& \kappa_{\rm eff}^2\,\iint_{\Omega}\,h_z^2\,
  \rho\,d\rho d\theta\nonumber \\
  &=& \kappa_{\rm eff}^2\,\sum_{i=1}^{N_\rho}\,\sum_{j=1}^{N_\theta}\,\int
  _{\theta_j}^{\theta_{j+1}}\,\int
  _{\rho_i}^{\rho_{i+1}}\,h_{z,i,j}^2\,\rho\,d\rho d\theta\nonumber \\
  &=&
  \kappa_{\rm eff}^2\,\sum_{i=1}^{N_\rho}\,\sum_{j=1}^{N_\theta}\,\frac{[{\rm Im}(1-L_{i,j})]^2}
  {a_\rho^2 \rho_{i+1/2}^2a_\theta^2}a_\rho\ \rho_{i+1/2}a_\theta\;.
\end{eqnarray}

\begin{figure*}
  \centering
  \includegraphics[scale=0.6]{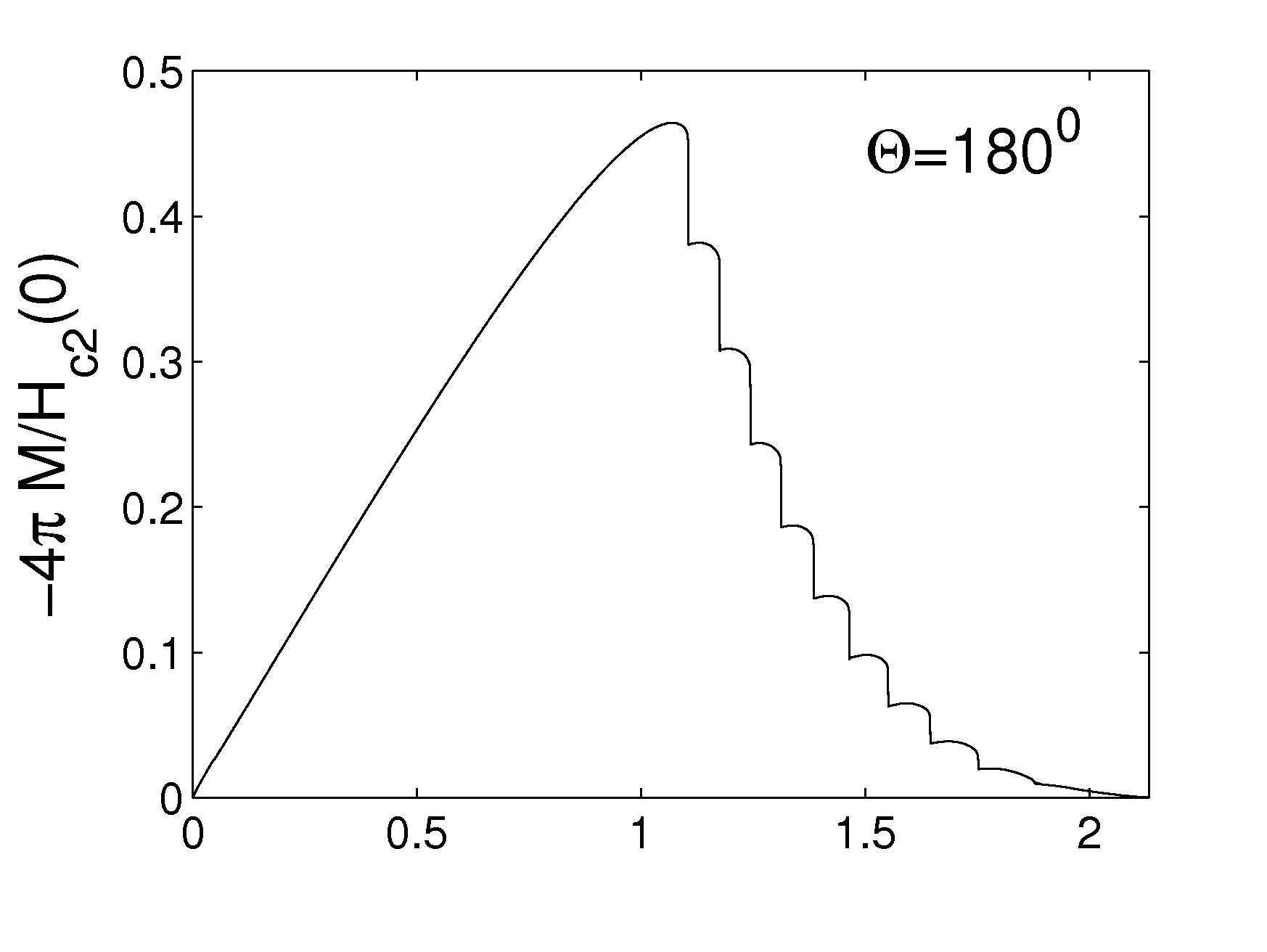}\hspace{-0.4cm}
  \includegraphics[scale=0.6]{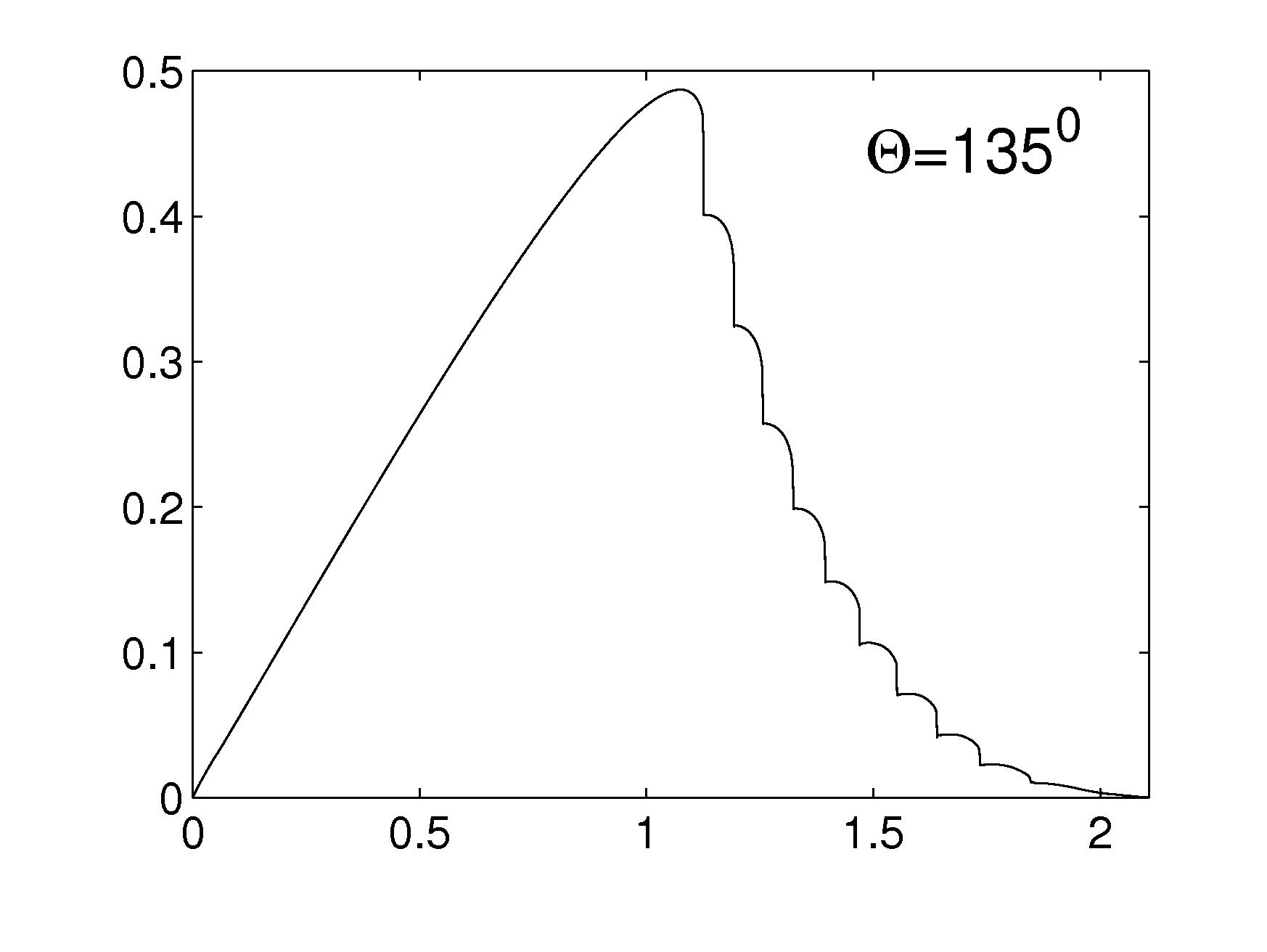}\\
  \vspace{-0.35cm}
  \includegraphics[scale=0.6]{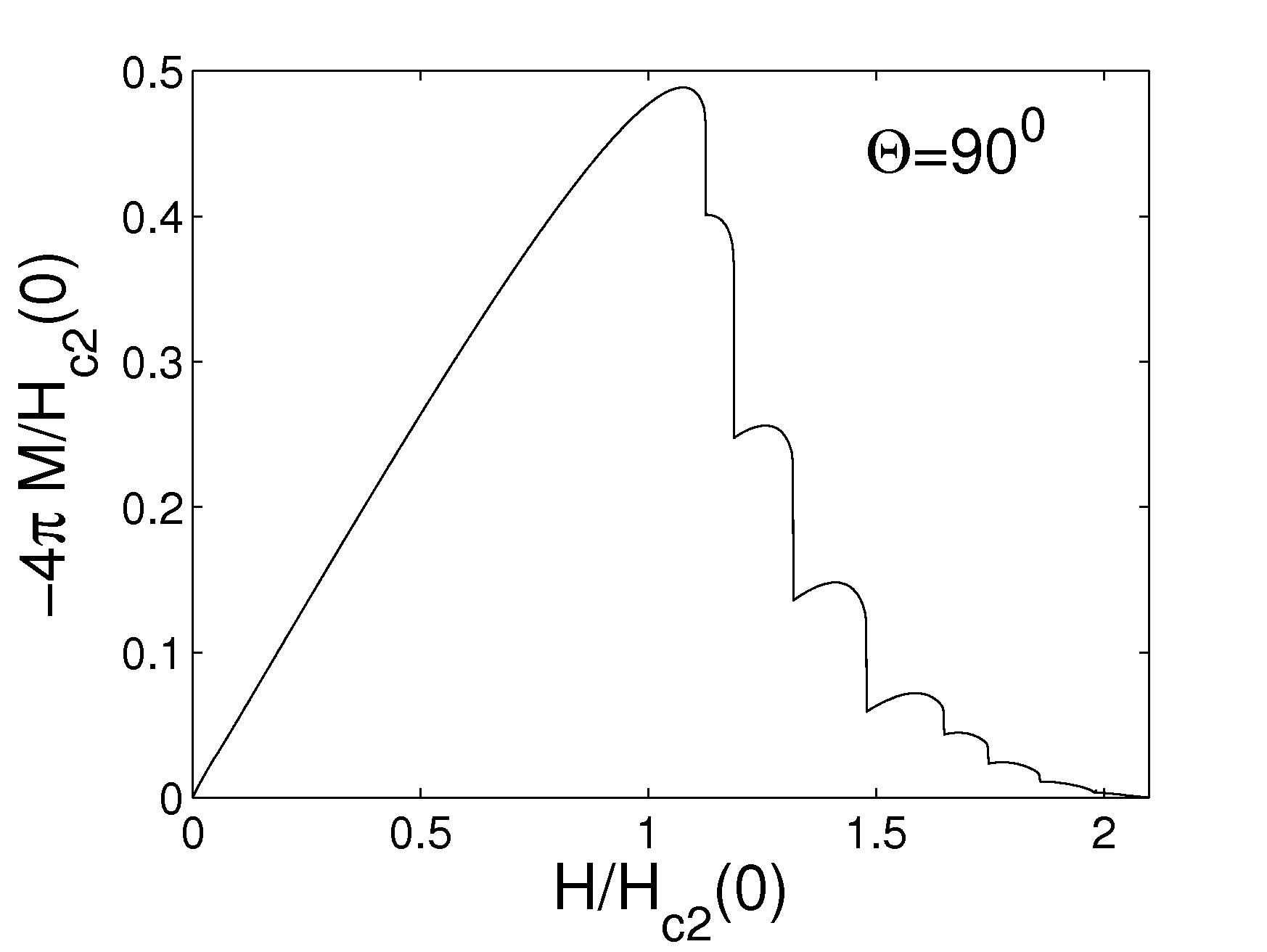}\hspace{-0.4cm}
  \includegraphics[scale=0.6]{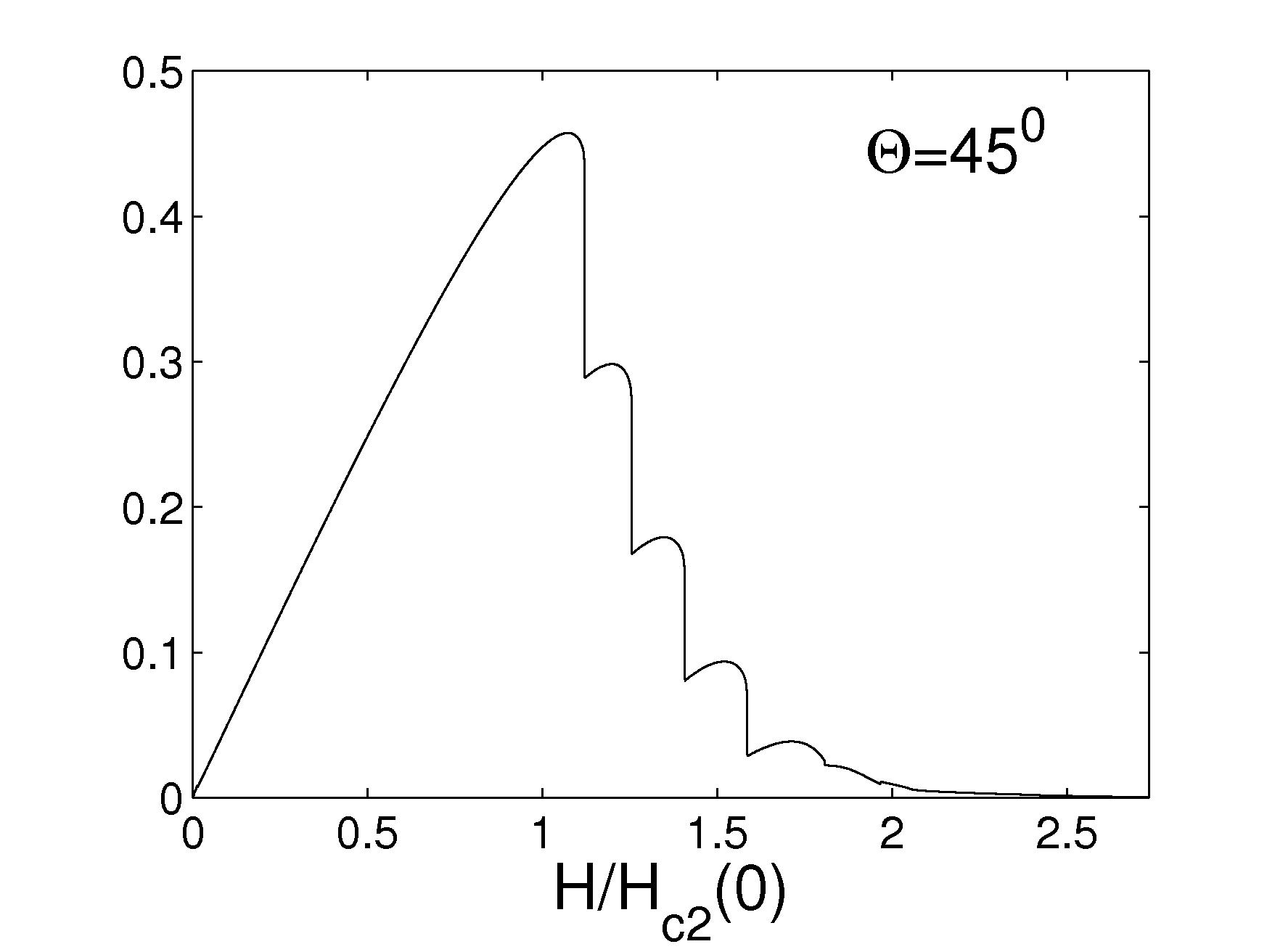}
  \caption{The magnetization curve as a function of the
  external applied magnetic field for four values of the angular
  width of the circular sector. Each jump in the magnetization indicates a phase transition. The
  corresponding configuration for each phase is indicated in Table~\ref{tab1}.}\label{fig2}
\end{figure*}

\begin{figure}
  \centering
  \includegraphics[scale=0.915]{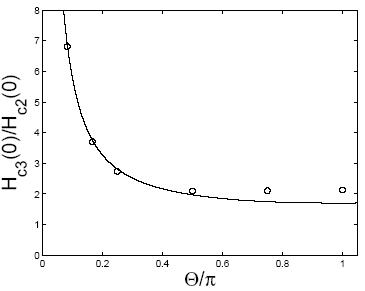}
  \caption{The nucleation field as function of the angular width of the circular sector. The solid
  line corresponds to equation (\ref{hc3}) taken from Ref.~\onlinecite{schweigert} and the open circles
  are the results found in
  the present simulation.}\label{fig3}
\end{figure}

\begin{figure*}
  \centering
  \includegraphics[scale=0.85]{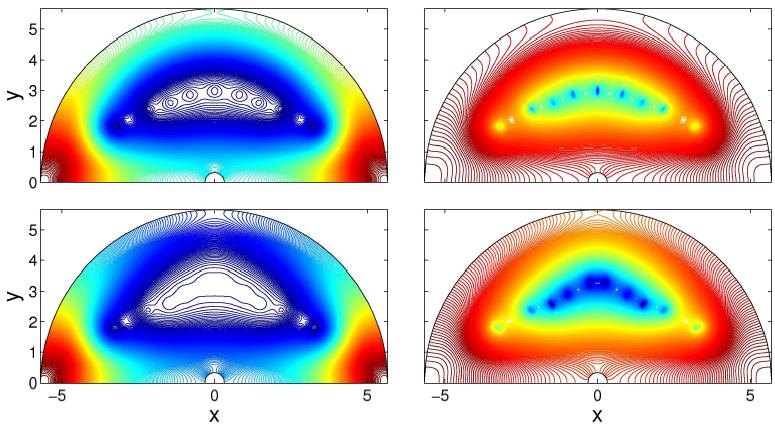}
  \caption{(Color online) Two dimensional density contour plots of $|\psi|$ for an angular
  width of $\Theta=180^0$ with $N=9$ and $N=10$ vortices, from the top to the bottom (left
  column; the right column corresponds to the same pictures, but in
  logarithm scale). Both pictures correspond to the
  stationary states as the vortices enter the sample. Notice that, once
  the equilibrium configuration is achieved, the vortices are symmetric with
  respect to the vertical axis. This feature is always present in the other
  geometries.}\label{fig4}
\end{figure*}

\begin{figure*}
  \centering
  \includegraphics[scale=0.85]{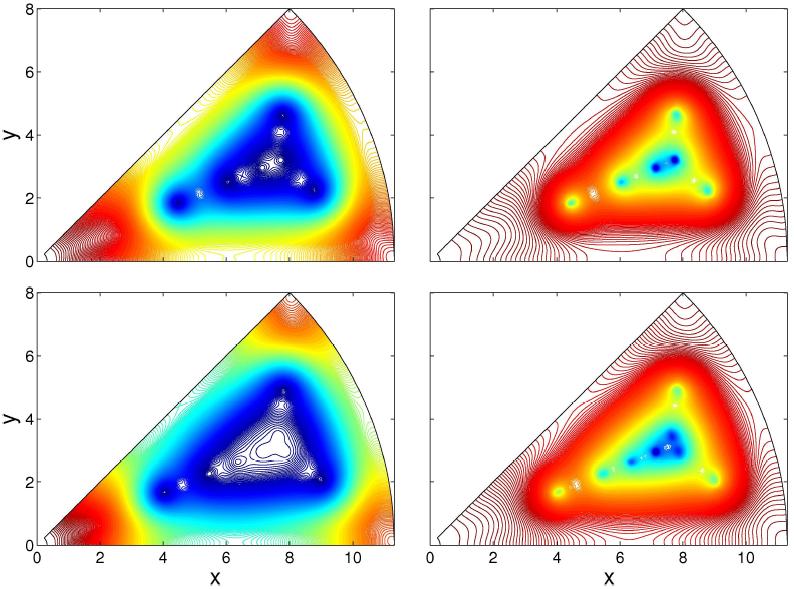}
  \caption{(Color online) The same as Fig.~\ref{fig4} for $\Theta=45^0$ with $N=6$ and $N=8$.}\label{fig5}
\end{figure*}

The total Helmholtz energy is then given by
$\mathscr{L}=\mathscr{L}_{\rm k}+\mathscr{L}_{\rm
c}+\mathscr{L}_{\rm f}$. The Gibbs free energy can be obtained by a
simple modification in the field energy. Instead of $h_z^2$ we would
have $(h_z-H)^2$, or $(h_{z,i,j}-H)^2$ in the discrete version.
Notice that the discrete TDGL equations could also be derived
through the following equations
\begin{eqnarray}
  (1-T)\frac{\partial\psi_{i,j}}{\partial t} &=& -\frac{1}{\mathscr{A}_i}
  \frac{\partial\mathscr{L}}{\partial\bar{\psi}_{i,j}}\,\nonumber \\
  \beta\frac{\partial A_{\alpha,i,j}}{\partial t} &=& -\frac{1}{2\mathscr{A}_{\alpha,i}}
  \frac{\partial\mathscr{L}}{\partial A_{\alpha,i,j}}\;,
\end{eqnarray}
where $\mathscr{A}_i=a_\rho\rho_ia_\theta$,
$\mathscr{A}_{\rho,i}=a_\rho\rho_{i+1/2}a_\theta$, and
$\mathscr{A}_{\theta,i}=a_\rho\rho_ia_\theta$, which are the areas
surrounded by the vertex and link points respectively. In order to
derive the discrete TDGL equations by this means, it is essential to
use the following relations
\begin{equation}
  \frac{\partial U_{\rho,i,j}}{\partial A_{\rho,i,j}} = -ia_\rho
  U_{\rho,i,j}\;,\;\;\;
  \frac{\partial U_{\theta,i,j}}{\partial A_{\theta,i,j}} =
  -i\rho_i a_\theta U_{\theta,i,j}\;,
\end{equation}
which can be easily shown from equations (\ref{linkvariables}).

The magnetization is $4\pi M=B-H$, where $B$ is the magnetic
induction which is given by the spatial average of the local
magnetic field. We have
\begin{equation}
    4\pi M=\frac{1}{\mathscr{A}}\sum_{i=1}^{N_\rho}\,\sum_{j=1}^{N_\theta}\,h_{z,i,j}
    \mathscr{A}_{\rho,i}-H\;,
\end{equation}
where $\mathscr{A}$ is total area of the circular sector.

The vorticity can be determined by integrating the phase $\varphi$
in each unit cell of the mesh. We have,
\begin{equation}
  N_{i,j} = \frac{1}{2\pi}\,\oint_{{\mathscr{C}}_{i,j}}\,\boldmath{\nabla}\varphi\cdot d{\bf r}\;,\;\;\;
  N = \sum_{i=1}^{N_\rho}\sum_{j=1}^{N_\theta}\,N_{i,j}\;,
\end{equation}
where $\mathscr{C}_{i,j}$ is a closed path with lower left and upper
right corner at $(i,j)$ and $(i+1,j+1)$, respectively.

In all of our numerical simulations described in the next Section,
we calculated $N$ in order to make sure that the number of vortices
agrees with what we see on the topological map of the order
parameter.

\section{Results and Discussion}\label{secRD}
The recurrence relations derived in the previous Section were
implemented as follows. We started from the Meissner state, where
$\psi=1$ and $U_\rho=U_\theta=1$ everywhere as the initial
condition. Then we let the time evolves until the system achieves
the stationary state.\cite{stationary} This is done by keeping the
external applied magnetic field $H$ constant. Next, we ramp up the
applied field by an amount of $\Delta H$. The stationary solution
for $H$ is then used as the initial state to determine the solution
for $H+\Delta H$, and so on. Usually we started from zero field and
increased $H$ until superconductivity is destroyed. As a criterion
for termination of the simulation, we monitored the Gibbs free
energy as a function of $H$. When the value of this quantity changes
its sign, then the transition from the superconducting to the normal
state sets in.

The parameters used in our numerical simulations were $\kappa=0.28$,
which is a typical value for Al;\cite{moshchalkov} the other
parameters are $d=0.1$, $T=0$, and $\beta=1$. The internal radio and
the area of the circular sector were taken fixed for any value of
angular width $\Theta$. We used $r=1/\pi$ and $16\pi$ for the area,
such that the external radio is given by
$R=\sqrt{32\pi/\Theta+r^2}$\,. The reason for taking these
parameters as such is because it makes possible comparison between
our results and previous ones (see Ref.~\onlinecite{baelus}, and
References therein). The size of the mesh varied according to the
value of $\Theta$. As a criterion we have taken the length of the
most external unit cell no larger then $0.25\times 0.25$. Since the
order parameter varied most significantly over a distance $\xi(T)$
(in real units), we are certain of not loosing this variation within
this criterion. We ramp up the applied magnetic field,  typically in
steps of $\Delta H=10^{-3}$.

In Fig.~\ref{fig2} we present the magnetization versus external
applied magnetic field curves for several values of the angular
width. These pictures present a typical profile of a magnetization
curve of a mesoscopic superconductor. It presents a series of
discontinuities, in which each jump signals the entrance of more
vortices into the sample. Notice that the lower critical field does
not vary with the shape of the circular sector. From this, the
immediate conclusion is that it depends only on the area but not on
$\Theta$. This is not an obvious result. In fact, in
Ref.~\onlinecite{baelus} numerical simulations were performed in
three different geometries: disk, square, and triangle and using the
same parameters as in the present contribution. They find that the
magnitude of the lower critical field is the same for the disk and
the square, but slightly larger for the triangle, despite all
geometries having the same area. Since we have the freedom to deform
the circular sector, it should be expected similar behavior for the
lower critical field, especially for small width angles $\Theta$.
However, the results shown in Fig.~\ref{fig2} do not seem to exhibit
this feature.

Another interesting feature present in the pictures of
Fig.~\ref{fig2} is that the SN transition field $H_{c3}(T)$ is
approximately the same for all angles greater than $90^0$. In
addition, $H_{c3}(T)$ is approximately the same for a disk and a
square of equivalent area. However, for smaller values of $\Theta$,
this critical field becomes significantly larger. Indeed, in
Ref.~\onlinecite{schweigert} $H_{c3}(T)$ was calculated numerically
for a wedge. The area they used for the wedge is $2.33$ larger than
the one we used here. They found that their results fit quite well
into the following expression
\begin{equation}\label{hc3}
    \frac{H_{c3}(T)}{H_{c2}(T)}=\frac{\sqrt{3}}{\Theta}\left ( 1+0.14804\Theta^2\frac{0.746\Theta^2}
    {\Theta^2+1.8794}\right )\;.
\end{equation}

In Fig.~\ref{fig3} we depict both the above expression and what we
have found for the nucleation field $H_{c3}(0)$ as a function of
$\Theta/\pi$. As can be seen from that Figure, for large angles, the
nucleation field is larger in the geometry we consider. This
suggests that, had we diminished the area of the circular sector,
this difference for large angles would have increased. Nonetheless,
for small angles all curves should collapse into a single curve,
which corresponds to the asymptotic behavior
$H_{c3}(T)/H_{c2}(T)=\sqrt{3}/\Theta$ determined in
Ref.~\onlinecite{schweigert}. This suggests that the superconductor
behaves as an unidimensional system as $\Theta$ becomes small, no
matter what the area is.

\begin{table}
\caption{The sequence of vortex configurations for four different
angles of the circular sector. The nomenclature used is explained in
the text. The configurations in brackets corresponds to what we
obtain not using a logarithm scale.}\label{tab1}
\begin{ruledtabular}
\begin{tabular}{rrrrr}
  $N$  & 180$^0$  & 135$^0$ & 90$^0$ & 45$^0$ \\
  \hline
  1 & 1S       & 1S         & 1S       & -       \\
  2 & 2S       & 2S         & -        & 2S      \\
  3 & 3S       & 3S         & 3S       & -       \\
  4 & 4S       & 4S         & -        & 4S      \\
  5 & 5S       & 5S         & 5S       & -       \\
  6 & 6S       & 6S         & -        & 6S      \\
  7 & 7S       & 7S         & 7S(5S1G$_2$)       & -       \\
  8 & 8S       & 8S(6S1G$_2$)         & 8S(5S1G$_3$)       & 8S(5S1G$_3$)      \\
  9 & 9S       & 9S(4S1G$_5$)         & 9S(5S1G$_4$)       & 9S(4S1G$_5$)      \\
  10& 10S(4S1G$_6$)& 10S(4S1G$_6$)   & 10S(3S1G$_7$)      & 10S(5S1G$_5$)     \\
  11& 11S(5S1G$_6$)      & 11S(2S1G$_9$)        & 11S(3S1G$_8$      & 11S(4S1G$_7$)     \\
\end{tabular}
\end{ruledtabular}
\end{table}

We also investigated the topology of the order parameter. Before
going any further, let us establish the criterion we use to
distinguish a single vortex from a giant vortex state. A giant
vortex is nucleated as two or more vortices collapse into a single
vortex in which all of them have, rigorously,  a common core center.
To describe an $N$ vortex state we use the following nomenclature.
We denote by $N_sS$, a multiple vortex configuration formed by $N_s$
single vortices. A single giant vortex of vorticity $N_g$ is denoted
by $1G_{N_g}$. For example, the $4S1G_2$ state is formed by four
single vortices and a double quantized giant vortex.

In all geometries we have considered, usually it occurs transitions
either from $N$ to $N+1$ or from $N$ to $N+2$ vortices. In
Fig.~\ref{fig4} and \ref{fig5} we depict $|\psi|$ for $\Theta=180^0$
and $\Theta=45^0$ respectively, and two stationary states with
different values of $H$. We have chosen transitions where we could
have the formation of a giant vortex. As can be seen in these
figures, we have the transitions $9S\rightarrow 4S1G_6$
($\Theta=180^0$) and $6S\rightarrow 5S1G_3$
($\Theta=45^0$).\cite{giant} However, if we look at the same
pictures in a logarithm scale, we still see that the core centers of
the vortices occupy different positions. So, within this criterion,
we cannot affirm that a giant vortex has been nucleated. For higher
vorticity, we have not observed any giant vortex either. A very
different scenario takes place in disks, squares, and triangles even
using the same parameters as in the present work.\cite{baelus}
Maybe, for smaller areas, giant vortex could be formed; we have not
tested this possibility. All possible configurations are summarized
in Table~\ref{tab1} up to $N=11$. Notice that the vortices are
always symmetrically distributed along the mediatrix.

\section{Summary}
In summary, an algorithm has been developed for solving the
Ginzburg-Landau theory for circular geometries. This will probably
make much easier to extend the $\psi U$ method for other geometries
in addition to the circular an rectangular ones. Furthermore, we
have applied the algorithm to circular sector and have found several
configurations for the vortex state in this geometry. Also, the
superconducting nucleation field has been evaluated. We have
presented some evidences that, as we diminishes the area of the
supercondutor, the nucleation field increases. However, as the
angular width goes to low values, this field exhibits an universal
behavior, independently of the area.

\begin{acknowledgments}
The authors thank the Brazilian Agencies FAPESP and CNPq for
financial support, and Dr.\ and Clecio C.\ de Souza Silva for useful
discussions.
\end{acknowledgments}


\begin{thebibliography}{99}
\bibitem{stationary} We use the follwoing criterion for the stationary state: if the highest difference
$||\psi(t)|-|\psi(t+\Delta t)||$, for any vertex point in the mesh,
is smaller than a certain precision $\epsilon$, then we go over the
next field. We have worked with a precision of $\epsilon=10^{-6}$
for $\Theta=45^0$ and $\epsilon=10^{-5}$ for the other angular
widths.
\bibitem{giant} We have used the hightest resolution as possible in order to
detect an individuality in the vortex configurations. Beyond some
critical resolution, the pictures do not present an change.
\end{thebibliography}
\end{document}